\documentclass[article,12pt,showkeys,showpacs] {revtex4}
\usepackage{amsfonts,graphicx,amsmath,longtable}
\begin{document}
\title{Steady-state mechanical squeezing in a double-cavity optomechanical system}
\author{Dong-Yang Wang}
\affiliation{Department of Physics, College of Science, Yanbian University, Yanji, Jilin 133002, People's Republic of China}
\author{Cheng-Hua Bai}
\affiliation{Department of Physics, College of Science, Yanbian University, Yanji, Jilin 133002, People's Republic of China}
\author{Hong-Fu Wang\footnote{E-mail: hfwang@ybu.edu.cn}}
\affiliation{Department of Physics, College of Science, Yanbian University, Yanji, Jilin 133002, People's Republic of China}
\author{Ai-Dong Zhu}
\affiliation{Department of Physics, College of Science, Yanbian University, Yanji, Jilin 133002, People's Republic of China}
\author{Shou Zhang}
\affiliation{Department of Physics, College of Science, Yanbian University, Yanji, Jilin 133002, People's Republic of China}

\begin{abstract}
We study the physical properties of double-cavity optomechanical system in which the mechanical resonator interacts with one of the coupled cavities and another cavity is used as an auxiliary cavity. The model can be expected to achieve the strong optomechanical coupling strength and overcome the optomechanical cavity decay, simultaneously. Through the coherent auxiliary cavity interferences, the steady-state squeezing of mechanical resonator can be generated in highly unresolved sideband regime. The validity of the scheme is assessed by numerical simulation and theoretical analysis of the steady-state variance of the mechanical displacement quadrature. The scheme provides a platform for the mechanical squeezing beyond the resolved sideband limit and addresses the restricted experimental bounds at present.
\end{abstract}
\maketitle
\section{Introduction}
The optomechanical system is a rapidly growing field in which researchers study the interaction between the optical and mechanical degrees of freedom via radiation pressure, optical gradient, or photothermal forces. Originally, the goal of studying the optomechanical interaction is to detect gravitational wave~\cite{CKRVRMP8052,BRPT9952}. As research continues, the optomechanical system has been developed to investigate quantum coherence for quantum information processing~\cite{FSPHY092,MPR04395} and quantum-to-classical transition studying in macroscopic solid-state devices~\cite{TKSCI08321,XFPRL9676}. Many projects of cavity optomechanics systems have been conceived and demonstrated experimentally, including red-sideband laser cooling in the resolved or unresolved sideband regime~\cite{INWTPRL0799,YYXQCPRA1591,XYPYPRA1592,YKWYPRA1490,FJASPRL0799,ZTMPRA1183}, coherent-state transiting between the cavity and mechanical resonator~\cite{LPRB1184,YAPRL12108}, normal-mode splitting~\cite{JITPRL08101,SKMMNAT09460}, quantum network~\cite{ZWLLPRA1591} backaction-evading measurements~\cite{AFKNJP0810,VYKSCI80209}, entanglement between mechanical resonator and cavity field or atom~\cite{DSAHPRL0798,CDPPRA0877,CDHAS1602}, macroscopic quantum superposition~\cite{JLPRL16116}, squeezing light~\cite{KCKMEP0979,TPRNPRX133,AFANJP1416,SMPRA1592}, and squeezing resonator~\cite{AJPRL09103,WGYPRA1388,HGPPRA1387,DCHASSR166,JYFPRA0979,MMPB00280,PAPPRB0470,AKJSPRA1082,XJLFPRA1591,AFAPRA1388,JCPRA1183}. In optomechanical systems, quantum fluctuations become the dominant mechanical driving force with strong radiation pressure, which leads to correlations between the mechanical motion and the quantum fluctuations of the cavity field~\cite{TRCSCI13339}.

In the above applications, quantum squeezing is important for studying the macroscopic quantum effects and the precision metrology of weak forces. The history of squeezing is linked intimately to quantum-limited displacement sensing~\cite{VF92}, and many schemes have been proposed to generate squeezing states in various systems~\cite{LHJHPRL8657,RLBJPRL8555,SPASPRA1388}. The squeezing of light is proposed for the first time using atomic sodium as a nonlinear medium~\cite{RLBJPRL8555}. In recent years, researchers have found that the optomechanical cavity, which can be regard as a low-noise Kerr nonlinear medium~\cite{CMSAPRA9449,SPPRA9449}, can be a better candidate to generate squeezing of the optical and mechanical modes. The squeezing of optical field is easy to be achieved in the optomechanical systems, and has been reported experimentally~\cite{TPRNPRX133,ASJJNAT13500,DTSTNAT12488}. Furthermore, many theoretical schemes have been proposed to generate mechanical squeezing in the optomechanical systems by using different methods~\cite{AJPRL09103,WGYPRA1388,HGPPRA1387,DCHASSR166,JYFPRA0979,MMPB00280,PAPPRB0470,AKJSPRA1082,XJLFPRA1591,AFAPRA1388,JCPRA1183}. For example, in 2010, Nunnenkamp \emph{et~al.}~\cite{AKJSPRA1082} proposed a scheme to generate mechanical squeezing via the quadratically nonlinear coupling between optical cavity mode and the displacement of a mechanical resonator. In 2011, Liao \emph{et~al.}~\cite{JCPRA1183} proposed a scheme to generate mechanical squeezing via periodically modulating the driving field amplitude at a frequency matching the frequency shift of the resonator. In 2013, Kronwald \emph{et~al.}~\cite{AFAPRA1388} proposed a scheme to generate mechanical squeezing by driving the optomechanical cavity with two controllable lasers with differing amplitudes in a dissipative mechanism. In 2015, L\"{u} \emph{et~al.}~\cite{XJLFPRA1591} proposed a scheme to generate steady-state mechanical squeezing via utilizing the mechanical intrinsic nonlinearity. With the deepening of research, the squeezing of mechanical mode has finally been observed experimentally by Wollman \emph{et~al.} in 2015~\cite{ECAJSCI15349} and attracted more and more attentions. While in these theoretical schemes, the mechanical resonator squeezing must rely on the resolved sideband limit, requiring a cavity decay rate smaller than the mechanical resonator frequency.

Traditionally and generally, the decay rate of cavity, which is a dissipative factor in optomechanical systems, is considered to have negative effect on the performance of quantum manipulation of mechanical modes. The optomechanical coupling strength $g=(\omega_{c}/L)\sqrt{\hbar/m\omega_{m}}$ (with the cavity frequency $\omega_{c}$, the mechanical resonator mass $m$, and the mechanical resonator frequency $\omega_{m}$) is inverse relation to the cavity length $L$. While the cavity quality factor $Q$ increases with increasing the cavity volume $V$. Thus it is difficult to achieve small decay rate and strong optomechanical coupling strength simultaneously. Here we propose a method to generate steady-state mechanical squeezing in a double-cavity optomechanical system with the highly dissipative cavity ($\kappa_{1}/\omega_{m}=100$). The scheme does not need to satisfy the conditions of the small cavity decay rate and the strong optomechanical coupling strength simultaneously. Different from the hybrid atom-optomechanical systems~\cite{CDPPRA0877,XYPYPRA1592,DCHASSR166}, the scheme does not have the challenge of putting a large number of atoms in the cavity. The coherently driving on the cavity mode is a monochromatic laser source which can generate strong optomechanical coupling between the mechanical and cavity modes. We show that, based on the mechanical nonlinearity and cavity cooling process in transformed frame, the steady-state mechanical squeezing can be successfully and effectively generated in the highly unresolved sideband regime via the coherent auxiliary cavity interfering. Unlike the dissipative coupling mechanism~\cite{AFANJP1416,AFAPRA1388,SPASPRA1388,SQHSPRA1592,SXHSSR144}, we utilize the destructive interference coming from the coherent auxiliary cavity to resist the influence of cavity decay.

The paper is organized as follows: In Section \uppercase\expandafter{\romannumeral2}, we describe the model of a double-cavity optomechanical system and derive the linearized Hamiltonian and the effective coupling between the auxiliary cavity and the mechanical resonator. In Section \uppercase\expandafter{\romannumeral3}, we engineer the mechanical squeezing and derive the analytical variance of the displacement quadrature of the mechanical resonator in the steady-state. In Section \uppercase\expandafter{\romannumeral4}, we study the relationship between the variance of mechanical mode and the system parameters and obtain the steady-state mechanical squeezing in the highly unresolved sideband regime by numerical simulations method. A conclusion is given in Section \uppercase\expandafter{\romannumeral5}.

\begin{figure}
  \includegraphics[width=4.5in]{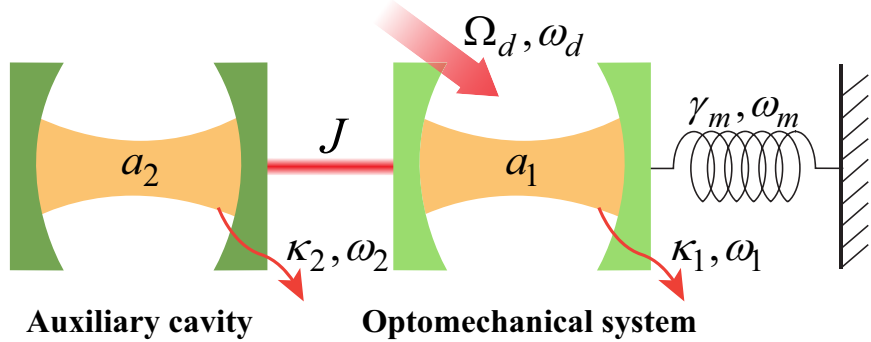}\\
  \caption{(Color online) Schematic diagram of a double-cavity optomechanical system. The cavity mode $a_{1}$ is coherently driven by an input laser with frequency $\omega_{d}$.}\label{f01}
\end{figure}

\section{System and Model}
We consider a double-cavity optomechanical system, which is composed of a mechanical resonator and two coupled single-mode cavities, depicted in Fig.~\ref{f01}. The mechanical resonator couples to the first dissipative cavity which is driven by an external laser field, forming the standard optomechanical subsystem. The second high \emph{Q} optical cavity is regarded as the auxiliary part, which couples to the first dissipative cavity with the coupling strength $J$. The total Hamiltonian $H=H_{0}+H_{\mathrm{I}}+H_{\mathrm{pump}}$, which describes the double-cavity optomechanical system, consists of three parts, which reads ($\hbar=1$), respectively,
\begin{eqnarray}\label{e01}
  H_{0}&=&\omega_{1}a_{1}^{\dag}a_{1}+\omega_{2}a_{2}^{\dag}a_{2}
        +\omega_{m}b^{\dag}b+\frac{\eta}{2}\left(b+b^{\dag}\right)^{4},\cr\cr
  H_{\mathrm{I}}&=&J\left(a_{1}^{\dag}a_{2}+a_{1}a_{2}^{\dag}\right)
                -ga_{1}^{\dag}a_{1}\left(b+b^{\dag}\right),\cr\cr
  H_{\mathrm{pump}}&=&\Omega_{d}\left(e^{-i\omega_{d}t}a_{1}^{\dag}
                    +e^{i\omega_{d}t}a_{1}\right).
\end{eqnarray}

The part $H_{0}$ accounts for the free Hamiltonian of the two cavity modes (with frequency $\omega_{1},~\omega_{2}$ and decay rate $\kappa_{1},~\kappa_{2}$, respectively) and the mechanical resonator (with frequency $\omega_{m}$ and damping rate $\gamma_{m}$). Here $a_{1}~(a_{1}^{\dag})$ is the bosonic annihilation (creation) operator of the first optical cavity mode, $a_{2}~(a_{2}^{\dag})$ is the bosonic annihilation (creation) operator of the second optical cavity mode, and $b~(b^{\dag})$ is the bosonic annihilation (creation) operator of the mechanical mode. The last term of $H_{0}$ describes the Duffing nonlinearity of the mechanical resonator with amplitude $\eta$. The intrinsic nonlinearity of the gigahertz mechanical resonator is usually very weak with nonlinear amplitude smaller than $10^{-15}\omega_{m}$. We can obtain a strong nonlinearity through coupling the mechanical mode to an auxiliary system~\cite{KAPRL09103,ASPRI03}, such as the nonlinear amplitude of $\eta=10^{-4}\omega_{m}$ can be obtained when we couple the mechanical resonator to an external qubit~\cite{XJLFPRA1591}.

The part $H_{\mathrm{I}}$ accounts for the interaction Hamiltonian consisting of the coupling interaction between two cavities and the optomechanical interaction derived from the radiation pressures. Where $J$ represents the intercavity coupling strength between cavity mode $a_{1}$ and $a_{2}$, and $g$ is the single-photon optomechanical coupling strength.

The part $H_{\mathrm{pump}}$ accounts for the external driving laser with frequency $\omega_{d}$ used to coherently pump the cavity mode $a_{1}$. The driving strength $\Omega_{\mathrm{d}}=\sqrt{2P\kappa_{1}/(\hbar\omega_{d}})$ is related to the input laser power $P$, the input laser frequency $\omega_{d}$, and the decay rate of cavity $\kappa_{1}$.

In the frame rotating at input laser frequency $\omega_{d}$, the Hamiltonian of the system is given by
\begin{eqnarray}\label{e02}
  H^{'}&=&-\delta_{1}a_{1}^{\dag}a_{1}-\delta_{2}a_{2}^{\dag}a_{2}+\omega_{m}b^{\dag}b
        +\frac{\eta}{2}\left(b+b^{\dag}\right)^{4}+J\left(a_{1}^{\dag}a_{2}
        +a_{1}a_{2}^{\dag}\right)\cr\cr
        &&-ga_{1}^{\dag}a_{1}\left(b+b^{\dag}\right)+\Omega_{d}\left(a_{1}
        +a_{1}^{\dag}\right),
\end{eqnarray}
where $\delta_{1}=\omega_{d}-\omega_{1}$ and $\delta_{2}=\omega_{d}-\omega_{2}$ are the detunings of the two cavity modes from the driving field, respectively. Considering the effect of the thermal environment, the quantum Heisenberg-Langevin equations for the system are written as
\begin{eqnarray}\label{e03}
    \dot{a_{1}}&=&\left(i\delta_{1}-\frac{\kappa_{1}}{2}\right)a_{1}-iJa_{2}
    +iga_{1}\left(b+b^{\dag}\right)-i\Omega_{d}-\sqrt{\kappa_{1}}a_{1in},\cr\cr
    \dot{a_{2}}&=&\left(i\delta_{2}-\frac{\kappa_{2}}{2}\right)a_{2}-iJa_{1}
    -\sqrt{\kappa_{2}}a_{2in},\cr\cr
    \dot{b}&=&\left(-i\omega_{m}-\frac{\gamma_{m}}{2}\right)b
    -2i\eta\left(b+b^{\dag}\right)^{3}+iga_{1}^{\dag}a_{1}-\sqrt{\gamma_{m}}b_{in},
\end{eqnarray}
where the corresponding noise operators $a_{1in},~a_{2in}$, and $b_{in}$ satisfy the following correlations:
\begin{eqnarray}\label{e04}
\langle a_{1in}(t)a_{1in}^{\dag}(t^{'})\rangle&=&\langle a_{2in}(t)a_{2in}^{\dag}(t^{'})\rangle~=~\delta(t-t^{'}),\cr\cr
\langle a_{1in}^{\dag}(t)a_{1in}(t^{'})\rangle&=&\langle a_{2in}^{\dag}(t)a_{2in}(t^{'})\rangle~=~0,\cr\cr
\langle b_{in}(t)b_{in}^{\dag}(t^{'})\rangle&=&(\bar{n}_{\mathrm{th}}+1)\delta(t-t^{'}),\cr\cr
\langle b_{in}^{\dag}(t)b_{in}(t^{'})\rangle&=&\bar{n}_{\mathrm{th}}\delta(t-t^{'}).
\end{eqnarray}
Here, $\bar{n}_{\mathrm{th}}=
\left\{\mathrm{exp}\left[\hbar\omega_{m}/(k_{B}T)\right]-1\right\}^{-1}$ is the mean thermal excitation number of bath of the mechanical resonator at temperature $T$, $k_{B}$ is the Boltzmann constant. And under the assumption of Markovian baths, the noise operators $a_{1in},~a_{2in}$, and $b_{in}$ have zero mean values.

\begin{figure}
  \includegraphics[width=4.5in]{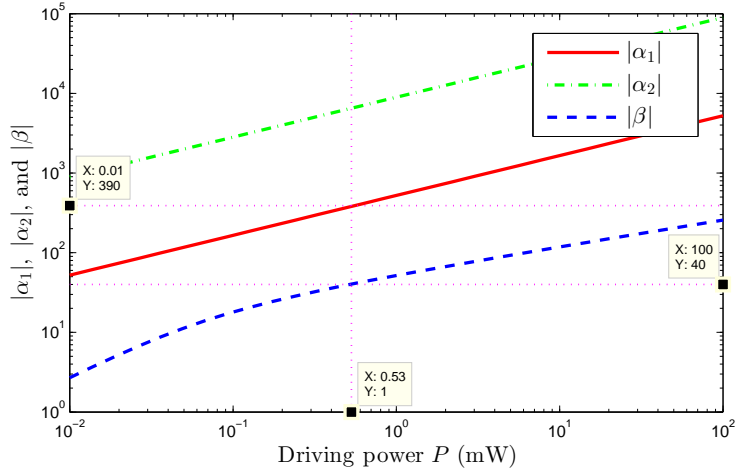}\\
  \caption{(Color online) The steady-state amplitudes $|\alpha_{1}|$, $|\alpha_{2}|$, and $|\beta|$ versus the driving power $P$. The parameters are chosen to be $\omega_{m}/(2\pi)=5~\mathrm{MHz}$, $\omega_{a}/(2\pi)=500~\mathrm{THz}$, $\delta_{1}=50\omega_{m}$, $\delta_{2}=1.05\omega_{m}$, $J=18\omega_{m}$, $g=10^{-3}\omega_{m}$, $\eta=10^{-4}\omega_{m}$, $\kappa_{1}=100\omega_{m}$, $\kappa_{2}=0.1\omega_{m}$, $\gamma_{m}=10^{-6}\omega_{m}$, and $\Omega_{d}=\sqrt{2P\kappa_{1}/(\hbar\omega_{d})}$.}\label{f02}
\end{figure}

Since the system is driven by a classical laser field, in the case of strong driving field, we can treat the field operators as the sum of their mean values and small quantum fluctuation. So we can apply a displacement transformation to linearize the equations, $a_{1}\rightarrow\alpha_{1}+a_{1},~a_{2}\rightarrow\alpha_{2}+a_{2},~b\rightarrow\beta+b$, where $\alpha_{1},~\alpha_{2}$, and $\beta$ are $c$ numbers denoting the mean values of the optical and mechanical modes. The mean values of the optical and mechanical modes satisfy the following equations:
\begin{eqnarray}\label{e05}
\dot{\alpha_{1}}&=&\left[i\left(\delta_{1}+g\beta+g\beta^{\ast}\right)
        -\frac{\kappa_{1}}{2}\right]\alpha_{1}-iJ\alpha_{2}-i\Omega_{d},\cr\cr
\dot{\alpha_{2}}&=&\left(i\delta_{2}-\frac{\kappa_{2}}{2}\right)\alpha_{2}
        -iJ\alpha_{1},\cr\cr
\dot{\beta}&=&\left(-i\omega_{m}-\frac{\gamma_{m}}{2}\right)\beta
-2i\eta\left(\beta^{\ast3}+\beta^{3}+3\beta^{\ast2}\beta
+3\beta^{\ast}\beta^{2}+3\beta^{\ast}+3\beta\right)+ig\alpha_{1}^{\ast}\alpha_{1}.
\end{eqnarray}
The steady-state amplitudes of the optical and mechanical modes are relative to the driving power $P$, and the relationship can be derived by solving the above equations under the condition of steady situation. One can see that when the driving power $P$ is in the microwatt range, the amplitudes of the cavity and mechanical modes satisfy the relationships: $|\alpha_{1}|,~|\beta|\gg1$, as shown in Fig.~\ref{f02}. And the amplitudes of the cavity and mechanical modes increase with increasing the driving power.  For example, at the point of the driving power $P=0.53~\mathrm{mW}$, the result of $|\alpha_{1}|\simeq390$ and $|\beta|\simeq40$ can be obtained, respectively.

Under the conditions of strong driving, the nonlinear terms are neglected. The quantum fluctuations satisfy the following linearized equations:
\begin{eqnarray}\label{e06}
\dot{a_{1}}&=&\left(i\Delta_{1}-\frac{\kappa_{1}}{2}\right)a_{1}-iJa_{2}
        +iG\left(b+b^{\dag}\right)-\sqrt{\kappa_{1}}a_{1in},\cr\cr
\dot{a_{2}}&=&\left(i\delta_{2}-\frac{\kappa_{2}}{2}\right)a_{2}-iJa_{1}
        -\sqrt{\kappa_{2}}a_{2in},\cr\cr
\dot{b}&=&\left(-i\tilde{\omega}_{m}-\frac{\gamma_{m}}{2}\right)b
    +iG\left(a_{1}+a_{1}^{\dag}\right)-2i\Lambda b^{\dag}-\sqrt{\gamma_{m}}b_{in},
\end{eqnarray}
with
\begin{align}\label{e07}
    \Delta_{1}&=\delta_{1}+2g|\beta|,&\tilde{\omega}_{m}&=\omega_{m}+2\Lambda,\cr\cr
    \Lambda&=3\eta\left(4|\beta|^{2}+1\right),&G&=g|\alpha_{1}|.
\end{align}
The linearized Hamiltonian is given by
\begin{eqnarray}\label{e08}
H_{\mathrm{L}}&=&-\Delta_{1}a_{1}^{\dag}a_{1}-\delta_{2}a_{2}^{\dag}a_{2}
        +\tilde{\omega}_{m}b^{\dag}b+\Lambda\left(b^{2}+b^{\dag2}\right)
        +J\left(a_{1}^{\dag}a_{2}+a_{1}a_{2}^{\dag}\right)\cr\cr
        &&-G\left(a_{1}+a_{1}^{\dag}\right)\left(b+b^{\dag}\right).
\end{eqnarray}
When considering the system-reservoir interaction, which results in the dissipations of the system, the full dynamics of the system is described by the master equation
\begin{eqnarray}\label{e09}
\dot{\rho}=-i\left[H_{\mathrm{\mathrm{L}}},\rho\right]+\kappa_{1}\mathcal{L}[a_{1}]\rho
    +\kappa_{2}\mathcal{L}[a_{2}]\rho+\gamma_{m}\left(\bar{n}_{\mathrm{th}}+1\right)\mathcal{L}[b]\rho
    +\gamma_{m}\bar{n}_{\mathrm{th}}\mathcal{L}[b^{\dag}]\rho,
\end{eqnarray}
where $\mathcal{L}[o]\rho=o\rho o^{\dag}-(o^{\dag}o\rho+\rho o^{\dag}o)/2$ is the standard Lindblad operators. $\kappa_{1}$, $\kappa_{2}$, and $\gamma_{m}$ are the decay rate of cavity mode $a_{1}$, $a_{2}$, and the damping rate of mechanical resonator, respectively. $\bar{n}_{\mathrm{th}}$ is the average phonon number in thermal equilibrium.

Since the decay rate of cavity 1 ($\kappa_{1}$) is much larger than the decay rate of cavity 2 ($\kappa_{2}$) and the damping rate of mechanical resonator ($\gamma_{m}$), the cavity mode $a_{1}$ can be eliminated adiabatically for the time scales longer than $\kappa_{1}^{-1}$. The steady solution of the first equation in Eq.~(\ref{e06}) about cavity mode $a_{1}$ can be written as
\begin{eqnarray}\label{e10}
    a_{1}=-\frac{iJ}{-i\Delta_{1}+\frac{\kappa_{1}}{2}}a_{2}
            +\frac{iG}{-i\Delta_{1}+\frac{\kappa_{1}}{2}}\left(b+b^{\dag}\right)
            -\frac{\sqrt{\kappa_{1}}}{-i\Delta_{1}+\frac{\kappa_{1}}{2}}a_{1in}.
\end{eqnarray}
Substituting Eq.~(\ref{e10}) into the rear two equations of Eq.~(\ref{e06}), we can obtain the effective coupling between the cavity mode $a_{2}$ and the mechanical mode $b$, which can be described by the following equations:
\begin{eqnarray}\label{e11}
    \dot{a_{2}}&=&\left(i\Delta_{\mathrm{eff}}-\frac{\kappa_{\mathrm{eff}}}{2}\right)a_{2}
            +iG_{\mathrm{eff}}\left(b+b^{\dag}\right)-A_{2in},\cr\cr
    \dot{b}&=&\left(-i\tilde{\omega}_{m}^{'}-\frac{\gamma_{m}}{2}\right)b
    +iG_{\mathrm{eff}}\left(a_{2}+a_{2}^{\dag}\right)-2i\Lambda^{'}b^{\dag}-B_{in},
\end{eqnarray}
where $A_{2in}$ and $B_{in}$ denote the modified noise terms, the effective parameters of the mechanical frequency, optomechanical coupling strength, detuning, decay rate, and coefficients of bilinear terms are given by
\begin{eqnarray}\label{e12}
  \tilde{\omega}_{m}^{'}&=&\tilde{\omega}_{m}
  +\frac{2G^{2}\Delta_{1}}{\Delta_{1}^{2}+\left(\frac{\kappa_{1}}{2}\right)^{2}}
  =\omega_{m}+2\Lambda^{'},\cr\cr
  G_{\mathrm{eff}}&=&\left|\frac{GJ}{\Delta_{1}\pm i\frac{\kappa_{1}}{2}}\right|,\cr\cr
  \Delta_{\mathrm{eff}}&=&\delta_{2}-\frac{J^{2}\Delta_{1}}{\Delta_{1}^{2}
  +\left(\frac{\kappa_{1}}{2}\right)^{2}},\cr\cr
  \kappa_{\mathrm{eff}}&=&\kappa_{2}+\frac{J^{2}\kappa_{1}}{\Delta_{1}^{2}
  +\left(\frac{\kappa_{1}}{2}\right)^{2}},\cr\cr
  \Lambda^{'}&=&\Lambda+\frac{G^{2}\Delta_{1}}{\Delta_{1}^{2}
  +\left(\frac{\kappa_{1}}{2}\right)^{2}}.
\end{eqnarray}
Thus the effective Hamiltonian is written as
\begin{eqnarray}\label{e13}
H_{\mathrm{eff}}=-\Delta_{\mathrm{eff}}a_{2}^{\dag}a_{2}+\tilde{\omega}_{m}^{'}b^{\dag}b
    -G_{\mathrm{eff}}\left(a_{2}+a_{2}^{\dag}\right)\left(b+b^{\dag}\right)
    +\Lambda^{'}\left(b^{2}+b^{\dag2}\right),
\end{eqnarray}
and the master equation becomes
\begin{eqnarray}\label{e14}
\dot{\rho}=-i\left[H_{\mathrm{eff}},\rho\right]+\kappa_{\mathrm{eff}}\mathcal{L}[a_{2}]\rho
    +\gamma_{m}\left(\bar{n}_{\mathrm{th}}+1\right)\mathcal{L}[b]\rho
    +\gamma_{m}\bar{n}_{\mathrm{th}}\mathcal{L}[b^{\dag}]\rho.
\end{eqnarray}
The effective Hamiltonian describes the effective interaction between the cavity 2 and the mechanical resonator. As we all know, if the Hamiltonian in the interaction picture has the form $b^{2}+b^{\dag2}$, the corresponding evolution operator is a squeezed operator.
\section{Engineering the mechanical squeezing}
Applying the unitary transformation $S(\zeta)=\mathrm{exp}[\zeta(b^{2}-b^{\dag2})/2]$, which is the single-mode squeezing operator with the squeezing parameter
\begin{eqnarray}\label{e15}
    \zeta=\frac{1}{4}\mathrm{ln}\left(1+\frac{4\Lambda^{'}}{\omega_{m}}\right),
\end{eqnarray}
to the total system. Then the transformed effective Hamiltonian becomes
\begin{eqnarray}\label{e16}
    H_{\mathrm{eff}}^{'}=S^{\dag}\left(\zeta\right)H_{\mathrm{eff}}S\left(\zeta\right)=
    -\Delta_{\mathrm{eff}}a_{2}^{\dag}a_{2}+\omega_{m}^{'}b^{\dag}b
    -G^{'}\left(a_{2}+a_{2}^{\dag}\right)\left(b+b^{\dag}\right),
\end{eqnarray}
with
\begin{eqnarray}\label{e17}
\omega_{\mathrm{m}}^{'}&=&\omega_{m}\sqrt{1+\frac{4\Lambda^{'}}{\omega_{m}}},\cr\cr G^{'}&=&G_{\mathrm{eff}}\left(1+\frac{4\Lambda^{'}}{\omega_{m}}\right)^{-\frac{1}{4}},
\end{eqnarray}
where $\omega_{\mathrm{m}}^{'}$ is the transformed effective mechanical frequency and $G^{'}$ is the transformed effective optomechanical coupling. The transformed Hamiltonian is a standard cavity cooling Hamiltonian and the best cooling in the transformed system is at the optimal detuning $\Delta_{\mathrm{eff}}=-\omega_{m}^{'}$. In the transformed frame, the master equation, which is used for describing the system-reservoir interaction, can be obtained via applying the squeezing transformation $S(\zeta)$ to the master equation Eq.~(\ref{e14}). The transformed master equation can achieve the cooling process. Here, $S^{\dag}(\zeta)\bar{n}_{\mathrm{th}} S(\zeta)=\bar{n}_{\mathrm{th}}^{'}=\bar{n}_{\mathrm{th}}\mathrm{cosh}(2\zeta)
+\mathrm{sinh}^{2}(\zeta)$ is the transformed thermal phonon number. The steady-state density matrix $\rho$ can be obtained by solving the master equation Eq.~(\ref{e14}). Defining the displacement quadrature $X=b+b^{\dag}$ for the mechanical mode, the steady-state variance of $X$ is given by $\langle\delta X^{2}\rangle=\langle X^{2}\rangle-\langle X\rangle^{2}$, which can be derived as
\begin{eqnarray}\label{e18}
    \langle\delta X^{2}\rangle=\left(2\bar{n}_{\mathrm{eff}}^{'}+1\right)e^{-2\zeta},
\end{eqnarray}
where $\bar{n}_{\mathrm{eff}}^{'}$ is the steady-state phonon number of the transformed system. When the best cooling in ideal situation $\bar{n}_{\mathrm{eff}}^{'}=0$ is achieved by the cooling process, the steady-state variance of the mechanical resonator displacement quadrature is $\langle\delta X^{2}\rangle=e^{-2\zeta}$.
\section{Numerical simulations and discussion}
\begin{figure}
  \includegraphics[width=4.5in]{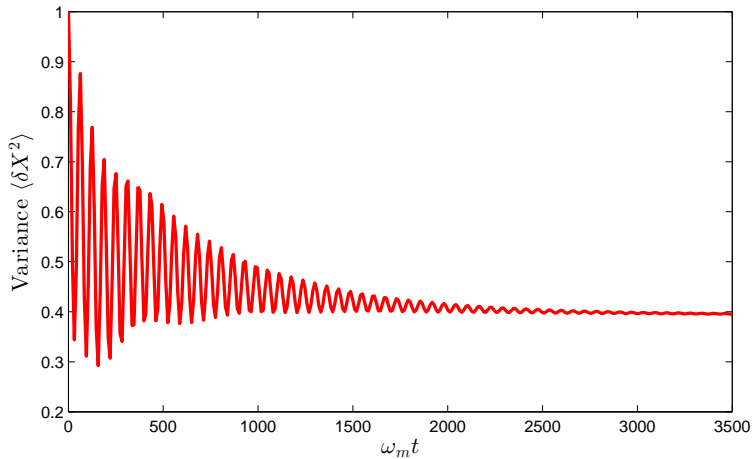}\\
  \caption{(Color online) The time evolution of variance $\langle\delta X^{2}\rangle$ about the mechanical displacement quadrature, and the other parameters are chosen to be the same as in Fig.~\ref{f02}.}\label{f03}
\end{figure}

In this section, we solve the original master equation Eq.~(\ref{e09}) numerically to calculate the steady-state variance of the mechanical displacement quadrature $X$. Firstly, we should provide the time evolution of variance $\langle\delta X^{2}\rangle$ about the mechanical displacement quadrature, which is shown in Fig.~\ref{f03}. It indicate that the variance $\langle\delta X^{2}\rangle$ gradually tends to be stable after a period of time. For simplicity, we have assumed that the system is initially prepared in its ground state and the system parameters are chosen to be the same as in Fig.~\ref{f02}.

\begin{figure}
  \includegraphics[width=4.5in]{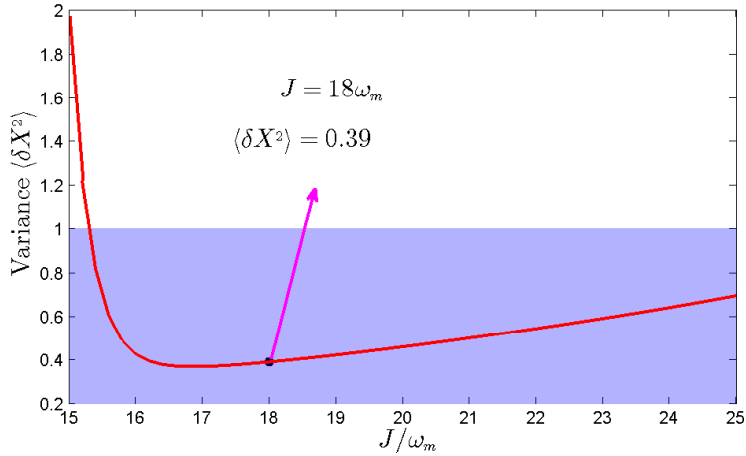}\\
  \caption{(Color online) The variance of the displacement quadrature $X$ relates to the intercavity coupling strength $J$ by solving the master equation Eq.~(\ref{e09}) numerically, and the other parameters are chosen to be the same as in Fig.~\ref{f02}.}\label{f04}
\end{figure}

\begin{figure}
  \includegraphics[width=4.5in]{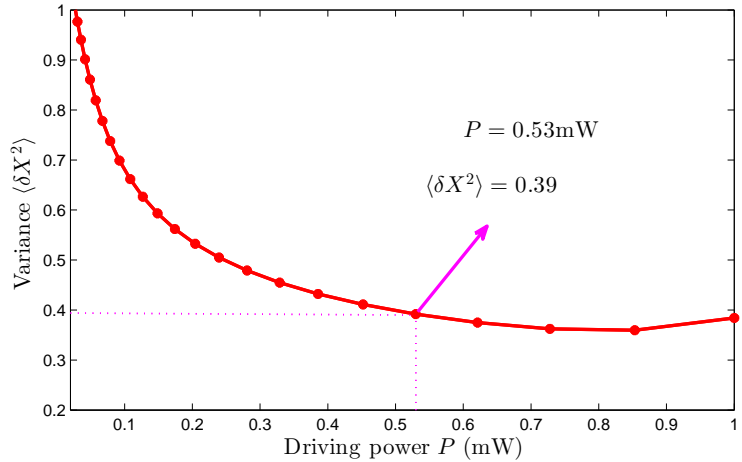}\\
  \caption{(Color online) The variance of the mechanical displacement quadrature $X$ relates to the driving power $P$ by solving the master equation numerically. The other parameters are chosen to be the same as in Fig.~\ref{f02}.}\label{f05}
\end{figure}

The relationship between the steady-state variance and intercavity coupling strength is shown in Fig.~\ref{f04}. Before we study their relationship, we should recalculate the steady-state amplitudes of the optical and mechanical modes $|\alpha_{1}|$, $|\alpha_{2}|$, and $|\beta|$ with the different intercavity coupling strengths. We can find that the steady-state mechanical squeezing can be achieved effectively when the intercavity coupling strength is appropriate, which reaches a balance between the enough large photons number in cavity 1 and the coherent auxiliary cavity interferences. However, when we remove the coherent auxiliary cavity interferences ($J=0$), the mechanical steady-state squeezing can not be obtained effectively under the present condition.
The relationship between the steady-state variance and driving power is shown in Fig.~\ref{f05}. One can see from Fig.~\ref{f04} that the steady-state squeezing of the mechanical resonator changes observably with the laser driving power. We can obtain the steady-state mechanical squeezing effectively when the driving power is in milliwatts level.

In the above, we study the steady-state squeezing of the mechanical resonator in a double-cavity optomechanical system and illustrate that the steady-state squeezing can be effectively generated in highly unresolved sideband regime with appropriate intercavity coupling strength and driving power. The experimental studies of the double-cavity optomechanical system with whispering-gallery microcavities have been reported~\cite{HSXJLFPRL14113,BSFFMGSNP142927,IHOKPRL10104,LXSCJLGNP14133}. When the decay rate of cavity is known, the maximum value of the squeezing parameter $\zeta$ is achieved at the point of $\Delta_{a}=\kappa_{1}/2$, which can be easily seen from Eq.~(\ref{e12}). Furthermore, the generated steady-state mechanical squeezing in the present scheme can be detected based on the method proposed in Refs.~\cite{XJLFPRA1591,DSAHPRL0798}. As illustrated in Refs.~\cite{XJLFPRA1591,DSAHPRL0798}, for detecting the steady-state mechanical squeezing, we can measure the position and the momentum quadratures of the mechanical resonator via homodyning detection of the output field of another auxiliary cavity mode with an appropriate phase, and the auxiliary cavity is driven by another pump laser field under a much weaker intracavity field so that its backaction on the mechanical mode can be neglected.

\section{Conclusions}
In conclusion, we have proposed a scheme for generating the steady-state squeezing of the mechanical resonator in a double-cavity optomechanical system via the mechanical nonlinearity and cavity cooling process in transformed frame. The steady-state squeezing of the mechanical resonator can be obtained in the highly unresolve sideband regime through the coherent auxiliary cavity interferences. Since the auxiliary cavity mode is not directly coupled to the mechanical resonator, it can be a high $Q$ optical cavity with big cavity volume $V$, while another cavity coupling with the mechanical resonator can have a short cavity length $L$ to possess good mechanical properties. The effective coupling between the mechanical resonator and the auxiliary cavity can be obtained by reducing the cavity mode adiabatically. We simulate the steady-state variance of the mechanical displacement quadrature numerically at a determinate laser driving power and find that under an appropriate intercavity coupling strength the steady-state mechanical squeezing can be achieved effectively in highly unresolve sideband regime. Our scheme opens up the possibility for application of cavity quantum optomechanics beyond the resolved sideband regime, addressing the restricted experimental bounds at present.
\begin{center}
{\bf{ACKNOWLEDGMENTS}}
\end{center}

This work was supported by the National Natural Science Foundation of China under
Grant Nos. 11264042, 11465020, 61465013, and 11564041.

\end{document}